\documentclass[a4paper]{jpconf}
\usepackage{graphicx}
\usepackage[caption=false]{subfig}
\begin{document}
\title{Multiplicity fluctuations at the quark-hadron phase transition from a fluid dynamical model}

\author{Christoph Herold$^1$, Marlene Nahrgang$^2$, Yupeng Yan$^1$, Chinorat Kobdaj$^1$}

\address{$^1$ School of Physics, Suranaree University of Technology, 111 University Avenue, Nakhon Ratchasima 30000, Thailand}
\address{$^2$ Department of Physics, Duke University, Durham, NC 27708, USA}

\ead{herold@g.sut.ac.th}

\begin{abstract}
The region of large net-baryon densities in the QCD phase diagram is expected to exhibit a first-order phase transition. Experimentally, its study 
will be one of the primary objectives for the upcoming FAIR accelerator. We model the transition between quarks and hadrons in a heavy-ion 
collision using a fluid which is coupled to the explicit dynamics of the chiral order parameter and a dilaton field. This allows us to investigate 
signals stemming from the nonequilibrium evolution during the expansion of the hot plasma. Special emphasis is put on an event-by-event analysis of 
baryon number fluctuations which have long since been claimed to be sensitive to a critical point.
\end{abstract}

\section{Introduction}
Although some aspects of the QCD phase structure have been revealed in the past years, still many things are unknown and require more thorough 
investigation. Regarding the chiral and deconfinement transition, we have learned from lattice QCD data that for small baryochemical potentials, 
the transition from a hadron gas to a quark-gluon plasma is a smooth crossover rather than a phase transition \cite{Aoki:2006we,Borsanyi:2010bp}. 
A critical endpoint (CEP) and first-order phase transition are expected from effective low-energy models \cite{Scavenius:2000qd,Schaefer:2007pw,Fukushima:2008wg,Herbst:2010rf}
and a Dyson-Schwinger approach \cite{Fischer:2014ata}, though any irrefutable proof is still missing. Experimentally, the dip in the recent measurement of 
directed flow as a function of beam energy by STAR \cite{Adamczyk:2014ipa} is sometimes argued to result from the presence of a phase transition. 
The measurement of proton number fluctuations during the beam energy scan program has revealed some non-monotonic behavior in higher moments 
such as the kurtosis \cite{Adamczyk:2013dal}. The idea 
hereby is that moments of fluctuations of conserved quantities are sensitive to a CEP as they scale with some powers of the correlation 
length \cite{Stephanov:2008qz}. However, these predictions rely on the assumption of thermodynamic equilibrium and to understand the observables we 
need to develop models which are able to describe the full nonequilibrium evolution of the hot and dense matter created in a heavy-ion collision. 
Effects like critical slowing down near a CEP \cite{Berdnikov:1999ph} and spinodal decomposition at a first-order phase transition 
\cite{Mishustin:1998eq,Randrup:2010ax} might influence the obtained signal. Our ansatz for such a model is to couple an ideal fluid of quarks and gluons to the explicit 
propagation of the relevant order parameters and additionally take into account friction and stochastic fluctuations \cite{Nahrgang:2011mg}. 
We have shown that this model successfully describes supercooling and critical slowing down \cite{Nahrgang:2011mv,Herold:2013bi} as well as 
domain formation due to spinodal dynamics \cite{Herold201414}. The goal of this work is to estimate the impact of nonequilibrium effects on an enhancement 
of net-baryon number fluctuations at a CEP and first-order phase transition \cite{Herold:2014zoa}.

\section{Nonequilibrium chiral fluid dynamics}
\label{sec:model}

We derive the coupled dynamics of order parameters and quark-gluon fluid from a linear sigma model with dilatons \cite{Sasaki:2011sd}
\begin{eqnarray}
\label{eq:Lagrangian}
{\cal L}&=&\overline{q}\left(i \gamma^\mu \partial_\mu-g_{\rm q} \sigma\right)q + \frac{1}{2}\left(\partial_\mu\sigma\right)^2 
+ \frac{1}{2}\left(\partial_\mu\chi\right)^2 + {\cal L}_A- U_{\sigma}-U_{\chi}~. \\
\end{eqnarray}
It describes the chiral dynamics of the light quarks $q=(u,d)$ coupled to the condensate $\sigma\sim\langle\bar q q\rangle$ which melts at 
high temperatures, thus restoring chiral symmetry. Additionally, gluons are incorporated with a constituent gluon field $A$ whose mass is generated 
by the glueball condensate $\langle\chi\rangle\sim\langle A_{\mu\nu}A^{\mu\nu}\rangle$, the so-called dilaton field. The quark-meson coupling 
$g_{\rm q}$ is fixed to a value of $3.37$ to reproduce the vacuum nucleon mass. Fixing the mass of the sigma meson at $900$~MeV results in a phase 
diagram with a CEP at temperature $T_{\rm CEP}=89$~MeV and quark chemical potential $\mu_{\rm CEP}=329$~MeV, with a first-order phase transition 
for larger chemical potentials. 

The mean-field effective thermodynamic potential is given by
\begin{equation}
 V_{\rm eff}=\Omega_{q\bar q}+\Omega_{A}+U_{\sigma}+U_{\chi}+\Omega_0~,
\end{equation}
with the quark and gluon contributions
\begin{eqnarray} 
\Omega_{\rm q\bar q}&=&-2 N_f N_c T\int\frac{\mathrm d^3 p}{(2\pi)^3} \left\{\ln\left[1+\mathrm e^{-\frac{E_{\rm q}-\mu}{T}}\right]+\ln\left[1+\mathrm e^{-\frac{E_{\rm q}+\mu}{T}}\right]\right\}~, \\
\Omega_{A}&=&2 (N_c^2-1) T\int\frac{\mathrm d^3 p}{(2\pi)^3} \left\{\ln\left[1-\mathrm e^{-\frac{E_A}{T}}\right]\right\}~,
\end{eqnarray}
Here, $E_{\rm q}=\sqrt{p^2+m_{\rm q}^2}$ and $E_A=\sqrt{p^2+m_A^2}$ denote the quasiparticle energies of constituent quarks and gluons, respectively. 
Note that we use the quark chemical potential $\mu=\mu_{\rm q}$. The constant $\Omega_0$ is added to set the total energy in vacuum to zero. The 
pressure of the ideal quark-gluon fluid is given by $p = -\Omega_{q\bar q}-\Omega_{A}$. 

For the dynamics of the sigma field we adopt the result from \cite{Nahrgang:2011mg} for our model, resulting in a Langevin equation containing 
a dissipative term $\eta_{\sigma}(T)\partial_t \sigma$ due to the possible decay of a sigma into a quark-antiquark pair and stochastic noise field
$\xi_{\sigma}$ resembling the back reaction of the fluid
\begin{equation}
\label{eq:eomsigma}
 \partial_\mu\partial^\mu\sigma+\eta_{\sigma}(T)\partial_t \sigma+\frac{\delta V_{\rm eff}}{\delta\sigma}=\xi_{\sigma}~.
\end{equation}
Due to kinematic reasons, no dissipative processes from interactions between quarks and the dilaton field are possible, therefore we use 
the classical Euler-Lagrange equation of motion for the dynamics of $\chi$,
\begin{equation}
\label{eq:eomchi}
 \partial_\mu\partial^\mu\chi+\frac{\delta V_{\rm eff}}{\delta\chi}=0~.
\end{equation}
Energy-momentum and net-baryon number are conserved through the equations
\begin{eqnarray}
\label{eq:fluidT}
\partial_\mu T^{\mu\nu}&=&-\partial_\mu\left(T_\sigma^{\mu\nu}+T_\chi^{\mu\nu}\right)~,\\
\label{eq:fluidN}
\partial_\mu N_{\rm q}^{\mu}&=&0~,
\end{eqnarray}
which together with the equation of state from the mean-field pressure determine the evolution of the fluid.
As can be seen, we allow for a direct transfer of energy-momentum from the order parameters to the quark-gluon fluid through the stochastic 
source terms $T_\sigma$ and $T_\chi$.

\section{Baryon number Fluctuations}
\label{sec:fluc}

\begin{figure}[t]
\centering
    \subfloat[\label{fig:suscep}]{
    \centering
    \includegraphics[scale=0.61,angle=270]{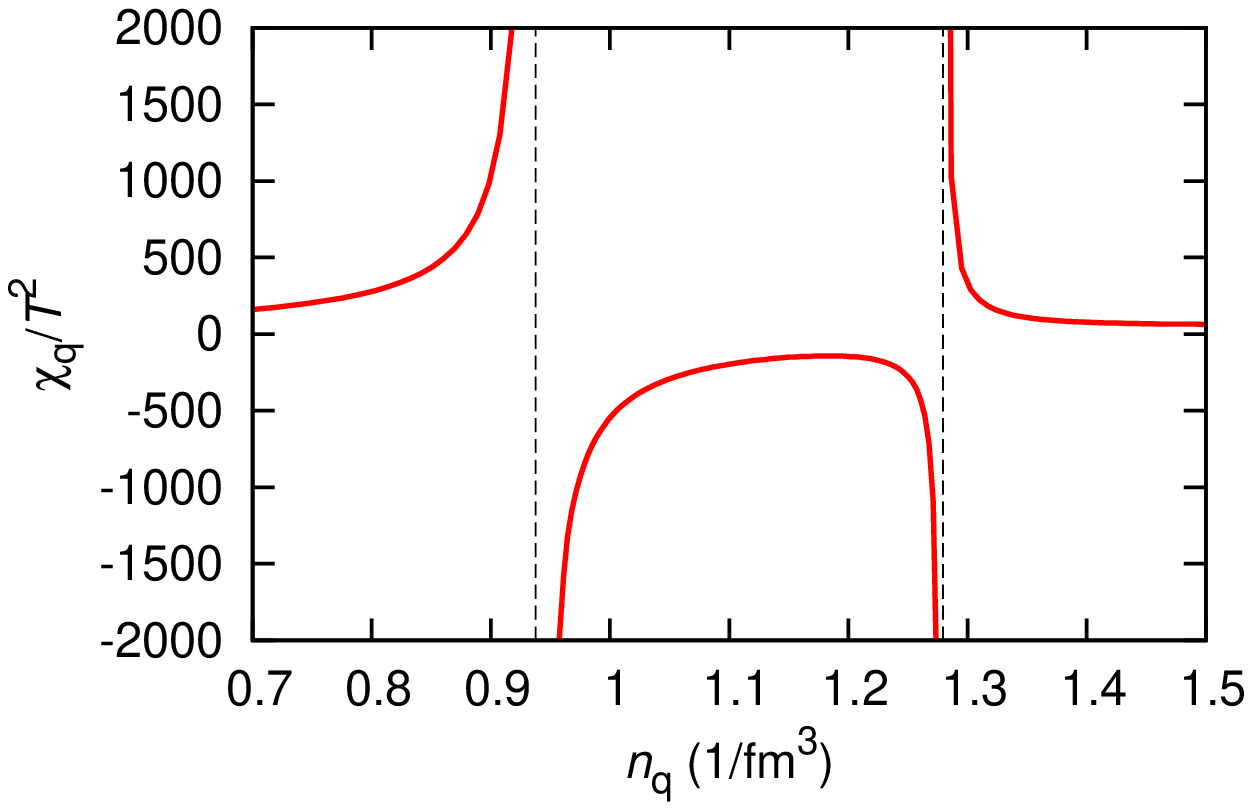}
    }
  \hfill
    \subfloat[\label{fig:kurtosis}]{
    \centering
    \includegraphics[scale=0.61,angle=270]{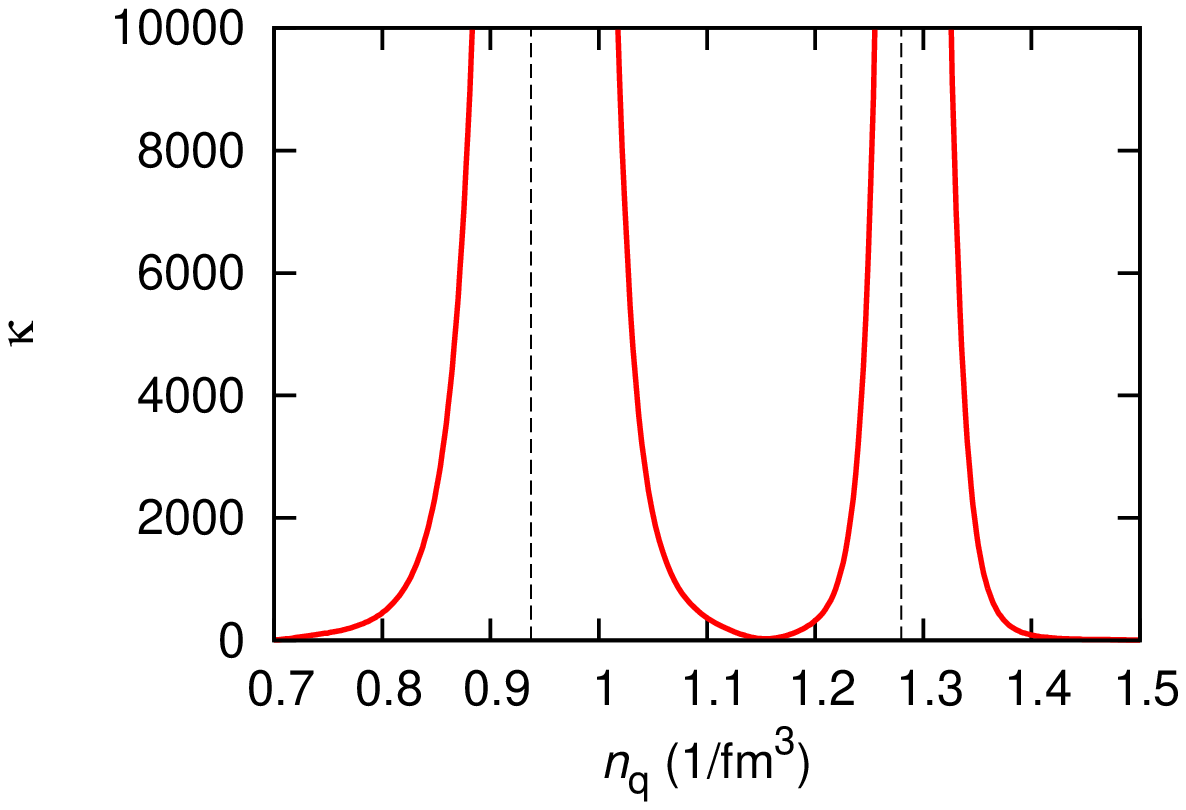}
    }
\caption[Quark number susceptibility and kurtosis]{Quark number susceptibility \subref{fig:suscep} and kurtosis \subref{fig:kurtosis} 
for a nonequilibrium first-order phase transition at $T=40$~MeV. Figure from \cite{Herold:2014zoa}.}
\label{fig:susceptibilities}
\end{figure}

Fluctuations of conserved quantities play an important role in identifying the location of a phase transition. They can be studied on several 
levels. For effective models, we may calculate for instance quark or baryon number susceptibilities according to the formula
\begin{equation}
 c_n = \frac{\partial^n(p/T^4)}{\partial(\mu/T)^n}~.
\end{equation}
as some derivative of the pressure with respect to the chemical potential. 
These coefficients are usually finite and diverge only at a CEP, thus marking a clear signal for a critical structure. However, as shown in 
\cite{Sasaki:2007db,Herold:2014zoa}, divergences occur also at the spinodal lines of a first-order phase transition if instabilities are taken 
into account. Fig. \ref{fig:susceptibilities} shows the susceptibility and kurtosis from the linear sigma model with dilatons for a constant 
temperature of $T=40$~MeV as a function of the net quark density. We see that both quantities diverge around the spinodal points, with critical 
indices of $2/3$ and $2$, respectively. This result is in agreement with what has been found for the NJL model in \cite{Sasaki:2007db}. 

\begin{figure}[t]
\centering
    \subfloat[\label{fig:ebesuscep}]{
    \centering
    \includegraphics[scale=0.72,angle=270]{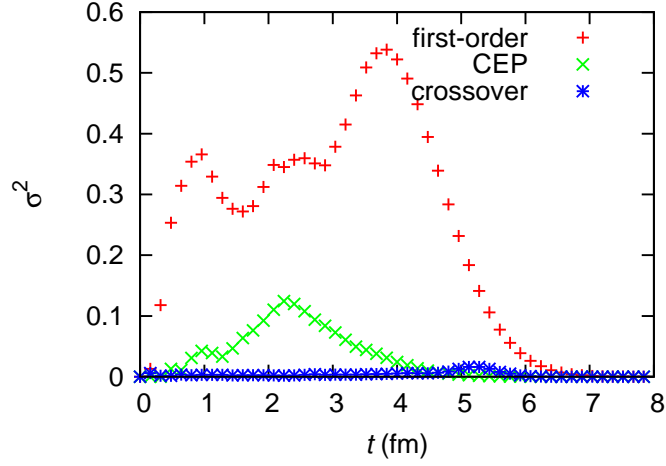}
    }
  \hfill
    \subfloat[\label{fig:ebekurtosis}]{
    \centering
    \includegraphics[scale=0.72,angle=270]{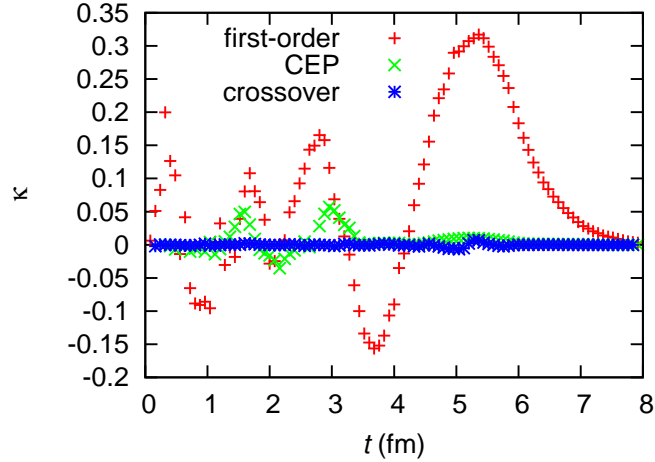}
    }
\caption[Event-by-event fluctuations]{Variance \subref{fig:ebesuscep} and kurtosis \subref{fig:ebekurtosis} of the net-baryon number from the fluid 
dynamical evolution as a function of time. Figure from \cite{Herold:2014zoa}.}
\label{fig:ebe}
\end{figure}

Given this, one may expect a clear enhancement of event-by-event fluctuations in experiment not only for a CEP but also, and probably even 
stronger, for a first-order phase transition. Note that the important aspect here is that the system is allowed to develop spinodal instabilities 
after the formation of a supercooled phase. Only in nonequilibrium, enhanced fluctuations can be observed for a first-order transition. 
We test this assumption by extracting fluctuations of the net-baryon number from dynamical simulations 
of heavy-ion collisions using the nonequilibrium chiral fluid dynamics model. For this purpose we begin with a spherical droplet of plasma, defined 
by an initial temperature and chemical potential. We choose these values such that the subsequent evolution proceeds through the desired region of 
the phase diagram, enabling us to study a crossover, CEP and first-order phase transition. As the total baryon number is conserved throughout the 
expansion of the hot and dense matter, we need to define some region of acceptance, as we are not able to study fluctuations in a grand canonical 
ensemble as the susceptibilities would indicate. We rather have to observe some appropriate part of the phase space in the canonical ensemble. 
As shown in \cite{Bzdak:2012an}, ratios of cumulants significantly depend on the ratio of measured to total baryons
$N/N_{\rm tot}$ due to the overall conservation law, making the choice of a suitable range of acceptance a delicate and nontrivial problem. We show results for the variance and kurtosis as a function of time in Fig. \ref{fig:ebe}.
The baryon number within a single event is calculated over all cells with rapidity $|y|<0.5$ and transverse momentum density 
$100 \mbox{ MeV/fm}^3<p_T<500 \mbox{ MeV/fm}^3$. In the upper plot we see the variance clearly enhanced for a CEP and even more 
for a first-order phase transition in comparison with a crossover scenario. The same holds for the kurtosis as shown in the lower plot. We see that 
it also becomes negative for both a CEP and a first-order phase transition. 

We also see that the enhanced fluctuations vanish for larger times, so the signal gets washed out in the hydrodynamic phase after passing the 
phase boundary. Therefore it is important to consider a criterion for a freeze-out to determine if the fluctuations in the density get finally 
imprinted in fluctuations of actual particle numbers as measured by a detector. We expect that due to baryon number conservation the fluctuations 
remain present even after particlization and final state interactions. However, this assumption has to be verified in the future.

\section*{Acknowledgements}

This work was funded by Suranaree University of Technology (SUT) and the CHE-NRU (NV.12/2557) project. The authors thank Igor Mishustin and 
Chihiro Sasaki 
for fruitful discussions and Dirk Rischke for providing the SHASTA code that was used numerical evolution of the ideal fluid. M. N. acknowledges support 
from the U.S. Department of Energy under grant DE-FG02-05ER41367 and a fellowship within the Postdoc-Program of the German 
Academic Exchange Service (DAAD).
The computing resources have been provided by the National e-Science Infrastructure 
Consortium of Thailand, the Center for Computer Services at SUT and the Frankfurt Center for Scientific Computing.

\section*{References}
\bibliographystyle{iopart-num}
\bibliography{mybib}

\end{document}